\PassOptionsToPackage{table}{xcolor}
\documentclass[sigconf]{acmart}
\usepackage{xcolor}

\usepackage{algorithm, algorithmic, amsmath, bm, subcaption, cleveref, multirow, makecell, mathtools, graphicx,xspace, booktabs}
\usepackage[normalem]{ulem}
\useunder{\uline}{\ul}{}

\newcommand{\modelname}{\textsf{CETNet}\xspace}

\AtBeginDocument{%
  }

\setcopyright{acmlicensed}
\copyrightyear{2018}
\acmYear{2018}
\acmDOI{XXXXXXX.XXXXXXX}

\acmConference[Conference acronym 'XX]{Make sure to enter the correct
  conference title from your rights confirmation email}{June 03--05,
  2018}{Woodstock, NY}
\acmISBN{978-1-4503-XXXX-X/18/06}

\newcommand{\beqa}{\begin{eqnarray}}
\newcommand{\eeqa}{\end{eqnarray}}
\newcommand{\beq}{\begin{equation}}
\newcommand{\eeq}{\end{equation}}
\newcommand{\ben}{\begin{enumerate}}
\newcommand{\een}{\end{enumerate}}
\newcommand{\bit}{\begin{itemize}}
\newcommand{\eit}{\end{itemize}}
\newcommand{\bi}{\begin{itemize} \item}
\newcommand{\ei}{\end{itemize}}

\newcommand{\begindef}{\begin{Definition} \rm}
\newcommand{\beginexa}{\begin{Example} \rm}
\newcommand{\beginthe}{\begin{Theorem} \rm}
\newcommand{\beginpro}{\begin{Proposition} \rm}
\newcommand{\beginlem}{\begin{Lemma} \rm}
\newcommand{\begincon}{\begin{Conjecture} \rm}
\newcommand{\begincor}{\begin{Corollary} \rm}

\newcommand{\eat}[1]{}

\def\papernumber #1 raised #2 {
\vspace{-#2}
\vbox to 0pt{\hfill\framebox{\bf Paper Number #1}}
\vspace{#2}
}

\begin{document}

\title{A Collaborative Ensemble Framework for CTR Prediction}


\author{Xiaolong Liu$^1$\quad Zhichen Zeng$^3$\quad Xiaoyi Liu$^2$\quad Siyang Yuan$^2$\quad Weinan Song$^2$\quad Mengyue Hang$^2$\quad Yiqun Liu$^2$\quad Chaofei Yang$^2$\quad Donghyun Kim$^2$\quad Wen-Yen Chen$^2$\quad Jiyan Yang$^2$\quad Yiping Han$^2$\quad Rong Jin$^2$\quad Bo Long$^2$\quad Hanghang Tong$^3$\quad Philip S. Yu$^1$\\
$^1$ University of Illinois Chicago, $^2$ Meta AI, $^3$ University of Illinois Urbana-Champaign  \\
\small\texttt{\{xliu262, psyu\}@uic.edu, \{zhichenz, htong\}@illinois.edu, \{xiaoyliu, syyuan, weinansong, hangm, yiqliu, yangcf10, donghyunk, wychen, chocjy, yipinghan, rongjinml, bolong\}\allowbreak @meta.com}}





\renewcommand{\shortauthors}{Trovato et al.}



\begin{abstract}
    Recent advances in foundation models have established scaling laws that enable the development of larger models to achieve enhanced performance, motivating extensive research into large-scale recommendation models.
    However, simply increasing the model size in recommendation systems, even with large amounts of data, does not always result in the expected performance improvements.
    In this paper, we propose a novel framework, \textbf{C}ollaborative \textbf{E}nsemble \textbf{T}raining \textbf{Net}work (\textbf{\modelname}), to leverage multiple distinct models, each with its own embedding table, to capture unique feature interaction patterns. 
    Unlike naive model scaling, our approach emphasizes diversity and collaboration through collaborative learning, where models iteratively refine their predictions. To dynamically balance contributions from each model, we introduce a confidence-based fusion mechanism using general softmax, where model confidence is computed via negation entropy. This design ensures that more confident models have a greater influence on the final prediction while benefiting from the complementary strengths of other models.
    We validate our framework on three public datasets (AmazonElectronics, TaobaoAds, and KuaiVideo) as well as a large-scale industrial dataset from Meta, demonstrating its superior performance over individual models and state-of-the-art baselines. Additionally, we conduct further experiments on the Criteo and Avazu datasets to compare our method with the multi-embedding paradigm. Our results show that our framework achieves comparable or better performance with smaller embedding sizes, offering a scalable and efficient solution for CTR prediction tasks.
    
\end{abstract}

\maketitle

\section{Introduction}\label{sec:intro}

Click-through rate (CTR) prediction is a critical task in many online services, such as e-commerce~\cite{commerce}, personalized recommendations~\cite{personalization1,personalization2,DLRM}, and digital advertising~\cite{ads1,ads2}. Accurately predicting user behavior allows platforms to deliver relevant content, improving both user experience and business outcomes. As data becomes more complex, designing models capable of capturing intricate feature interactions has become increasingly important.

Recent advances in CTR prediction have leveraged deep learning methods to effectively model complex patterns in user-item interactions~\cite{wide&deep,pan2024ads}. State-of-the-art approaches focus on designing sophisticated feature embedding techniques to represent categorical data, employing interaction layers to capture higher-order correlations among features~\cite{dcn,dcnv2,deepfm,xdeepfm}, and integrating attention mechanisms to emphasize important user behaviors~\cite{tin,DIN,DIEN,DMIN}. Furthermore, recent research has explored multi-task learning~\cite{mult-task1,pan2024ads} and multi-modal fusion~\cite{multi-modal1,multi-modal2} to enhance prediction accuracy by integrating various signals. These advancements aim to capture delicate relationships and improve the overall effectiveness of CTR models in personalized recommendation settings.

Traditional CTR models typically use a single embedding table for categorical features, limiting their ability to capture diverse interaction patterns. Recent work~\cite{multi-emb} shows that simply increasing model and embedding sizes does not guarantee better performance due to \textit{embedding collapse}, where the embedding matrix becomes nearly low-rank and occupies a low-dimensional subspace, restricting the model’s ability to capture information.
A promising direction is to adopt a multi-embedding approach, where multiple models—or the same model with different initializations or configurations—use distinct embedding tables. This allows each model to focus on different aspects of feature interactions, uncovering complementary patterns that may not be captured by a single embedding. By combining the strengths of these diverse embeddings, the overall representation becomes more expressive, improving the model’s ability to predict user behavior accurately.

The simple multi-embedding paradigm, while effective in capturing diverse feature interactions, faces several challenges. First, using multiple embedding tables without coordination can lead to redundant or conflicting representations, limiting the overall expressiveness of the model. Second, it is difficult to determine how much each embedding should contribute to the final prediction, as naive aggregation methods, such as concatenation or averaging, do not account for the varying confidence levels of different models. Third, without proper alignment, multiple models trained independently within a multi-embedding framework can result in uneven learning, where some models dominate while others underperform~\cite{dominance}, reducing the effectiveness of the ensemble.


In this paper, we propose \modelname to integrate two Meta's models, InterFormer~\cite{interformer} and DHEN~\cite{DHEN}, to leverage their complementary strengths to improve CTR prediction in internal on-line deployment.
Our framework offers several key advantages:
(1) Leveraging complementary strengths: By ensembling InterFormer and DHEN with distinct embedding tables, we capture both sequential and hierarchical feature interactions, enhancing the overall expressiveness of the feature representation.
(2) Dynamic contribution balancing: Our confidence-based fusion mechanism addresses the limitations of naive aggregation by dynamically weighting each model’s contribution based on its confidence, ensuring that more reliable models exert greater influence on the final prediction.
(3) Aligned joint learning: To prevent uneven learning and model dominance, we introduce collaborative learning with symmetric KL divergence, which aligns the predictions of the models and ensures that both contribute meaningfully to the ensemble, improving synergy and overall performance.
We validate the effectiveness of our framework through extensive experiments on three public datasets—Amazon, TaobaoAds, and KuaiVideo—demonstrating that our ensemble framework consistently outperforms individual models and other baselines. 
We also conduct additional experiments on the Criteo and Avazu datasets, where our method proves superior to the multi-embedding paradigm while using smaller embeddings.

In summary, the key contributions of this paper are as follows: 
\begin{itemize}
    \item We propose a novel ensemble framework that integrates multiple models with distinct embedding tables, effectively capturing complementary feature interactions to enhance CTR prediction. 
    \item Our framework integrates collaborative learning through KL divergence to align the predictions of different models, enhancing synergy and ensuring more effective learning.
    \item We introduce a confidence-based fusion mechanism that dynamically adjusts each model’s contribution based on its prediction confidence, leading to more robust predictions. 
    \item Our framework demonstrates superior performance across both public datasets as well as in internal deployment. It consistently outperforms state-of-the-art baselines and multi-embedding approaches with a more efficient design. 
\end{itemize}

The remainder of the paper is organized as follows: Section~\ref{sec:related} reviews related work. Section~\ref{sec:method} introduces our ensemble framework, including the confidence-based fusion and collaborative learning components. Section~\ref{sec:exp} describes the experimental setup and evaluation. Finally, Section~\ref{sec:con} concludes the paper.

\section{Related Works}\label{sec:related}
\subsection{Click-Through-Rate Prediction}
In this section, we provide an overview of the most representative models of CTR prediction from feature interaction models and user behavior models.
Feature interaction models focus on learning relationships between features to improve prediction performance. Traditional models such as Factorization Machines (FM)~\cite{rendle2010factorization} model pairwise feature interactions efficiently, but they struggle to capture high-order interactions. Building on FM, DeepFM~\cite{deepfm} integrates a neural network to learn both low-order and high-order feature interactions in a unified framework.
Wide \& Deep~\cite{wide&deep} combines linear (wide) and deep components to handle both memorization and generalization patterns.
DCNv2~\cite{dcnv2} extends the original Deep \& Cross Network~\cite{dcn} by introducing advanced cross layers that iteratively learn feature interactions.
More recently, DHEN~\cite{DHEN} leverages multiple heterogeneous interaction modules in a hierarchical ensemble structure to effectively capture complex, high-order feature interactions.
User behavior models focus on capturing sequential interactions from a user’s historical behavior to better predict future clicks. 
DIN~\cite{DIN} introduces an attention mechanism to select relevant parts of the user’s behavior sequence for each target item, modeling evolving interests. 
DIEN~\cite{DIEN} introduces a GRU-based structure to further capture sequential dependencies and evolving user interests over time. 
DMIN~\cite{DMIN} disentangles multiple underlying user preferences within behavior sequences. 

In recent research, the multi-embedding paradigm has gained significant attention. 
The work of~\cite{multi-emb} addresses the issue of embedding collapse, where embeddings occupy low-dimensional subspaces, limiting the model’s scalability. 
It proposes a multi-embedding design that assigns multiple embedding tables with interaction-specific modules, enabling the model to learn diverse patterns and mitigate collapse. 
A further study~\cite{pan2024ads} explores encoding features with inherent priors and introduces practical methods to mitigate dimensional collapse and disentangle user interests.



\subsection{Collaborative Learning}
Collaborative learning has emerged as an effective strategy in deep learning, particularly in scenarios where multiple models need to enhance each other’s performance. 
A prominent example is Deep Mutual Learning~\cite{DML}, where multiple models are trained simultaneously, each acting as a student network that learns not only from the ground truth but also from the predictions of other student networks through knowledge distillation~\cite{hinton2015distilling}. 
Codistillation~\cite{codistillation} improves distributed training by enabling parallel models to align their predictions through periodic prediction exchanges.
ONE~\cite{ONE} builds multiple branch classifiers and uses a gate controller to align their predictions, enabling real-time knowledge distillation.
KDCL~\cite{KDCL} trains multiple student networks with different capacities collaboratively learn from each other via online knowledge distillation.

Recently, collaborative learning has been applied to CTR prediction tasks. The work by~\cite{zhu2020ensembled} proposes using an ensemble of teacher models to transfer knowledge to a student model. Similarly,~\cite{yilmaz2024mutual} explores mutual learning for CTR prediction, where multiple models are trained collaboratively, learning from each other to improve performance. However, unlike computer vision (CV), CTR models heavily rely on embedding tables. Simply training multiple branches on a shared embedding table can lead to suboptimal results due to conflicts in learning. The multi-embedding paradigm offers a promising solution to overcome this challenge by assigning separate embedding tables to different models, allowing them to learn complementary interaction patterns.

\section{Methodology}\label{sec:method}
\begin{figure*}[htbp]
    \centering
    \includegraphics[scale=0.68]{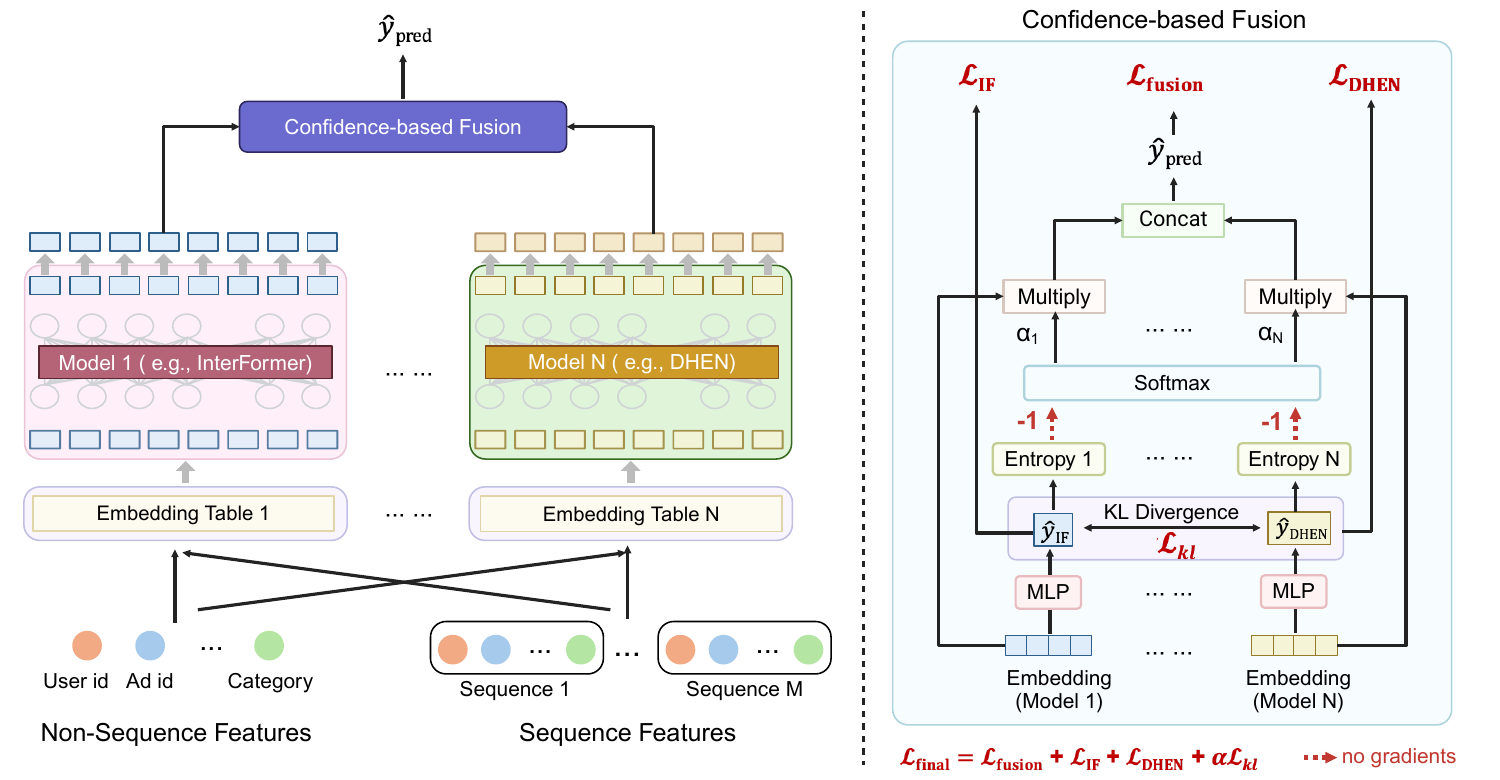}
    \caption{The framework of \modelname. 
    The left figure illustrates the overall framework of our ensemble with N models (for simplicity, we depict two models—InterFormer and DHEN—as used in our main experiments). 
    Each model utilizes its own distinct embedding table. The output embeddings from these models are combined to leverage the unique strengths of each model.
    The right figure shows the confidence-based fusion module, where each model produces embeddings and confidence scores.
    In addition to the prediction loss on the final output $\hat{\text{y}}_{pred}$, individual prediction losses are also applied to the model outputs $\hat{\text{y}}_{IF}$ and $\hat{\text{y}}_{DHEN}$.
    A KL divergence loss is introduced to facilitate collaborative learning between the models.
    The dashed arrow indicates stopped gradient flow.
    }
    \label{fig:model}
\end{figure*}
In this section, we first formulate the CTR prediction problem and then introduce our proposed framework, \modelname, as shown in Figure~\ref{fig:model}. Although our experiments focus on ensembling two of Meta's models, InterFormer and DHEN, the framework is general and can extend to ensemble $N$ models. 
The key components include multi-embedding paradigm~\ref{sec:multi-emb}, a collaborative learning mechanism~\ref{sec:cl}, and a softmax-based confidence fusion module~\ref{sec:fusion}.

\subsection{Problem Formulation}
The goal of CTR prediction task is to estimate the probability that a user will click on a given item.
Formally, let $\mathcal{U}$ denote the set of users and $\mathcal{I}$ denote the set of items.
For each user $u \in \mathcal{U}$ and item $i \in \mathcal{I}$, the objective is to learn a function $\hat{\text{y}} = f(u, i, x)$ that predicts $\hat{\text{y}}$, the probability of a click, where $x$ represents the associated features.
The input features \( x \) consist of three types: \textbf{sparse features} \( x_{\text{sparse}} \), \textbf{dense features} \( x_{\text{dense}} \), and \textbf{sequence features} \( x_{\text{seq}}^u \). 
Sparse features are categorical data such as user ID, item ID, and item category, which are mapped to dense vectors via embedding tables, represented as \( x_{\text{sparse}} = \{x_{\text{id}}^u, x_{\text{id}}^i, x_{\text{cat}}^i, \dots\} \). 
Dense features are continuous values, such as a user’s age or an item’s price, used either directly or after normalization, represented as \( x_{\text{dense}} = \{x_{\text{age}}^u, x_{\text{price}}^i, \dots\} \). 
Sequence features capture users' historical interactions with items, represented as \( S^u = \{s_1^u, s_2^u, \dots, s_N^u\} \), where \( u \) denotes the specific user. 
The sequence is truncated or padded to a fixed length \( N \) to ensure consistent input size, keeping only the most recent interactions if necessary or adding padding if the sequence is shorter.
In real-world scenarios, users often generate multiple interaction sequences. We denote these sequences for a given user \( u \) as \( S^u_1, \dots, S^u_k \), where \( k \) is the total number of sequences associated with the user.
Binary cross-entropy loss is the objective function for optimization.

\subsection{Multi-Embedding Paradigm}\label{sec:multi-emb}
Traditional CTR models typically use a single embedding table to transform categorical features into dense vector representations. However, relying on a single embedding table may limit the ability to capture the complex and diverse feature interaction patterns present in heterogeneous data. To address these limitations, previous work~\cite{multi-emb} proposed the multi-embedding paradigm, which mitigates embedding collapse and improves scalability by leveraging multiple embeddings for enhanced feature interaction learning.
Inspired by this paradigm, our framework employs two specialized models: DHEN and InterFormer, each equipped with its own embedding table. These models capture complementary feature interactions and ensure that the overall model benefits from diverse interaction patterns:

\begin{itemize}
    \item \textbf{DHEN}~\cite{DHEN} (Deep Hierarchical Ensemble Network) leverages multiple heterogeneous interaction modules to learn a hierarchy of interactions across different orders. Its hierarchical structure ensures that both low-order and high-order interactions are effectively captured, resulting in a more expressive feature representation.
    \item \textbf{InterFormer}~\cite{interformer} conducts deep interaction learning across heterogeneous features (including dense, sparse, and sequence features). It selects relevant information effectively without relying on aggressive aggregation, preserving the quality of critical feature interactions.
\end{itemize}

By assigning each model its own independent embedding table, the framework ensures that both hierarchical and deep feature interactions are captured, enriching the overall expressiveness of the learned feature representations.
Each model maintains a distinct embedding table: DHEN employs \( \mathbf{E}_{\text{DHEN}} \), while InterFormer utilizes \( \mathbf{E}_{\text{IF}} \). The sparse feature embeddings are retrieved through embedding lookups as follows:

\begin{align}
    \mathbf{e}_{\text{sparse}}^{\text{DHEN}} = \mathbf{E}_{\text{DHEN}}[x_{\text{sparse}}], \quad 
    \mathbf{e}_{\text{sparse}}^{\text{IF}} = \mathbf{E}_{\text{IF}}[x_{\text{sparse}}],\label{eq:sparse}
\end{align}
where \( x_{\text{sparse}} \) contains the indices of the sparse features used to retrieve the corresponding rows.
$\mathbf{e}_{\text{sparse}}^{\text{DHEN}}, \mathbf{e}_{\text{sparse}}^{\text{IF}} \in \mathbb{R}^{n \times d}$, where $n$ denotes the number of sparse features and $d$ is the embedding size.
For sequence features, each model retrieves a sequence of embeddings from its respective embedding table, while dense features are transformed using separate MLPs for each model:
\begin{align}
    \centering
    &\mathbf{S}^{\text{DHEN}} = [\mathbf{E}_{\text{DHEN}}[S_1^u]\| \dots \| \mathbf{E}_{\text{DHEN}}[S_N^u]],
    \mathbf{e}^{\text{DHEN}}_{\text{dense}} = \text{MLP}_{\text{DHEN}}(x_{\text{dense}}), \\   
    &\mathbf{S}^{\text{IF}} = [\mathbf{E}_{\text{IF}}[S_1^u]\| \dots \| \mathbf{E}_{\text{IF}}[S_N^u]], \quad 
    \mathbf{e}^{\text{IF}}_{\text{dense}} = \text{MLP}_{\text{IF}}(x_{\text{dense}}). \label{eq:sequence}
\end{align}
where $\mathbf{S}^{\text{DHEN}}, \mathbf{S}^{\text{IF}} \in \mathbb{R}^{k \times N \times d}$ and $\mathbf{e}^{\text{DHEN}}_{\text{dense}}, \mathbf{e}^{\text{IF}}_{\text{dense}} \in \mathbb{R}^{d}$. 
The outputs of InterFormer and DHEN are generated by feeding the input features into each model:

\begin{align}
    \mathbf{e}_{\text{IF}} = \text{InterFormer}(\mathbf{e}_{\text{sparse}}^{\text{IF}}, \mathbf{S}^{\text{IF}}, \mathbf{e}_{\text{dense}}^{\text{IF}}), \\
    \mathbf{e}_{\text{DHEN}} = \text{DHEN}(\mathbf{e}_{\text{sparse}}^{\text{DHEN}}, \mathbf{S}^{\text{DHEN}}, \mathbf{e}_{\text{dense}}^{\text{DHEN}}).
\end{align}

By maintaining separate embedding tables and MLPs for each model, InterFormer and DHEN capture distinct yet complementary feature interactions. InterFormer is designed to emphasize sequential learning, effectively modeling temporal dependencies and patterns in user behavior. On the other hand, DHEN focuses on capturing complex, high-order, and hierarchical feature interactions, making it well-suited for identifying intricate relationships among features. This complementary design ensures that both models contribute unique strengths, enhancing the overall performance of the ensemble.

\subsection{Collaborative Learning}\label{sec:cl}
\subsubsection{Challenge.}
In traditional ensemble models, individual models may dominate the learning process, leading to imbalanced performance~\cite{dominance}. 
To address this, we adopt a collaborative learning approach, where InterFormer and DHEN are trained in parallel, each focusing on different feature interaction patterns. InterFormer specializes in processing heterogeneous features such as dense, sparse, and sequential data, while DHEN captures high-order feature interactions through its hierarchical structure. 

One challenge in collaborative learning is ensuring that both models contribute meaningfully to the final prediction. Without proper coordination, one model could dominate, leading to imbalanced learning. To address this, we adopt a symmetric KL divergence term that aligns the predicted distributions of both models, encouraging mutual refinement without sacrificing their individual strengths.
The symmetric KL divergence for a dataset with \( N \) samples is the sum of the KL divergence in both directions:

\[
\mathcal{L}_{kl} = \frac{1}{2} \sum_{i=1}^{N} \mathcal{L}_{\text{KL}}^{(i)}(\hat{\text{y}}_{\text{IF}}, \hat{\text{y}}_{\text{DHEN}}) 
+ \frac{1}{2} \sum_{i=1}^{N} \mathcal{L}_{\text{KL}}^{(i)}(\hat{\text{y}}_{\text{DHEN}}, \hat{\text{y}}_{\text{IF}}),
\]
where the individual KL divergences for a sample \( i \) are given by:

\[
\mathcal{L}_{\text{KL}}^{(i)}(\hat{\text{y}}_{\text{IF}}, \hat{\text{y}}_{\text{DHEN}}) = \hat{\text{y}}_{\text{IF}}^{(i)} \log \frac{\hat{\text{y}}_{\text{IF}}^{(i)}}{\hat{\text{y}}_{\text{DHEN}}^{(i)}} 
+ (1 - \hat{\text{y}}_{\text{IF}}^{(i)}) \log \frac{1 - \hat{\text{y}}_{\text{IF}}^{(i)}}{1 - \hat{\text{y}}_{\text{DHEN}}^{(i)}},
\]

\[
\mathcal{L}_{\text{KL}}^{(i)}(\hat{\text{y}}_{\text{DHEN}}, \hat{\text{y}}_{\text{IF}}) = \hat{\text{y}}_{\text{DHEN}}^{(i)} \log \frac{\hat{\text{y}}_{\text{DHEN}}^{(i)}}{\hat{\text{y}}_{\text{IF}}^{(i)}} 
+ (1 - \hat{\text{y}}_{\text{DHEN}}^{(i)}) \log \frac{1 - \hat{\text{y}}_{\text{DHEN}}^{(i)}}{1 - \hat{\text{y}}_{\text{IF}}^{(i)}}.
\]
This refinement term encourages consistency between the two models while allowing them to focus on different aspects of the feature interactions. 

\subsubsection{Collaboration vs. Diversity.} Our framework leverages both collaborative learning and diversity to enhance ensemble performance. While collaboration and diversity might seem contradictory, they are complementary in our approach. The collaborative learning component, implemented with symmetric KL divergence, ensures balanced contributions from each model, preventing any single model from becoming overly dominant.

Diversity, in this context, refers to the variation in feature representations produced by the different models in the ensemble. By using distinct embedding tables and targeting different interaction patterns, sequential for InterFormer and hierarchical for DHEN, each model captures unique aspects of the data, adding diverse perspectives to the final prediction.
Together, these mechanisms enable our framework to achieve robust predictions by balancing aligned learning goals with diverse feature representations.

\subsection{Confidence-Based Fusion}\label{sec:fusion}

Once the two models—InterFormer and DHEN—generate their respective embeddings, we combine them using a confidence-based fusion mechanism inspired by~\cite{han2022multimodal, zhang2024multimodal}. This mechanism ensures that each model’s contribution is proportional to its confidence in the prediction. We use the negation entropy of each model’s predicted probability distribution to measure its confidence. Lower entropy indicates higher certainty, and thus a larger contribution to the final embedding.
The entropy for each model is computed as:

\[
H(P) = - \hat{y} \text{log} \hat{y} - (1 - \hat{y})\text{log}(1 - \hat{y}), 
\]
where $\hat{y}$ is the prediction.
Let \( P_{\text{IF}} \) and \( P_{\text{DHEN}} \) be the predicted probability distributions from InterFormer and DHEN, respectively.
The inverse entropy for each model, which serves as the confidence score, is given by:

\[
C_{\text{IF}} = - H(P_{\text{IF}}), \quad C_{\text{DHEN}} = - H(P_{\text{DHEN}}).
\]

To ensure that the confidence values do not interfere with the gradient flow during back-propagation, we detach the entropy values from the computation graph. This prevents the confidence scores from affecting the updates to each model’s individual projection layer (MLP) applied to its output embedding.

\subsubsection{Softmax Weight Calculation and Embedding Fusion}

We apply the softmax function to the inverse entropy values to generate the fusion weights for the embeddings:

\[
w_{\text{IF}} = \frac{\exp(C_{\text{IF}})}{\exp(C_{\text{IF}}) + \exp(C_{\text{DHEN}})}, \quad
w_{\text{DHEN}} = \frac{\exp(C_{\text{DHEN}})}{\exp(C_{\text{IF}}) + \exp(C_{\text{DHEN}})}.
\]

Using these weights, the final fused embedding \( \mathbf{e}_{\text{fused}} \) is computed via weighted concatenation as:

\[
\mathbf{e}_{\text{fused}} = [w_{\text{IF}} \cdot \mathbf{e}_{\text{IF}} \| w_{\text{DHEN}} \cdot \mathbf{e}_{\text{DHEN}}],
\]
where \( \mathbf{e}_{\text{IF}} \) and \( \mathbf{e}_{\text{DHEN}} \) are the embeddings generated by InterFormer and DHEN, respectively.

\subsubsection{Final Prediction}

The fused embedding \( \mathbf{e}_{\text{fused}} \) is used to generate the final prediction through a readout function:

\[
\hat{\text{y}}_{\text{fused}} = \sigma(\mathbf{W} \cdot \mathbf{e}_{\text{fused}} + b),
\]
where \( \sigma(\cdot) \) is the sigmoid activation function, \( \mathbf{W} \) is the weight matrix, and \( b \) is the bias term. This final prediction, \( \hat{\text{y}}_{\text{fused}} \), represents the likelihood of a click event.

\subsubsection{Benefits of Confidence-Based Fusion}
The confidence-based fusion mechanism ensures that the model with higher prediction certainty has a greater influence on the final output. This adaptive strategy enhances prediction robustness by dynamically leveraging each model’s strengths for individual instances. Additionally, detaching the entropy values stabilizes learning by preventing unintended gradient flow through the confidence scores.

\subsubsection{Objective Function}

Our framework optimizes three key binary cross-entropy losses: one for the output of the DHEN model, one for the InterFormer model, and one for the fused output. We define a unified loss function as:

\[
\mathcal{L}(\hat{\text{y}}_i) = - \frac{1}{N} \sum_{i=1}^{N} \left( \text{y}_i \log(\hat{\text{y}}_i) + (1 - \text{y}_i) \log(1 - \hat{\text{y}}_i) \right),
\]
where \( \text{y}_i \) is the ground truth label, \( \hat{\text{y}}_i \) is the predicted probability for the \( i \)-th sample, and \( N \) is the total number of samples.
The specific losses for each component are expressed as:

\[
\mathcal{L}_{\text{DHEN}} = \mathcal{L}(\hat{\text{y}}_{\text{DHEN}}), \quad
\mathcal{L}_{\text{IF}} = \mathcal{L}(\hat{\text{y}}_{\text{IF}}), \quad
\mathcal{L}_{\text{fusion}} = \mathcal{L}(\hat{\text{y}}_{\text{fused}}),
\]
where \( \textbf{e}_{\text{DHEN}}, \textbf{e}_{\text{IF}}, \) and \( \textbf{e}_{\text{fused}} \) are the embeddings from the DHEN model, InterFormer model, and the fused output, respectively.
The final objective function is defined as:

\[
\mathcal{L}_{\text{final}} = \mathcal{L}_{\text{fusion}} + \mathcal{L}_{\text{DHEN}} + \mathcal{L}_{\text{IF}} + \alpha \mathcal{L}_{kl},
\]
where \( \alpha \) is a hyperparameter controlling the impact of the KL divergence term.
This objective function ensures that all three outputs—DHEN, InterFormer, and the fused model—are optimized simultaneously. The use of symmetric KL divergence aligns the individual models' predictions, enhancing collaboration and improving overall performance.

\section{Experiment}\label{sec:exp}

\subsection{Experiment Setup}\label{sec:exp-setup}
\subsubsection{Datasets.} 
Our method is evaluated on three benchmark real-world datasets, including AmazonElectronics~\cite{he2016ups}, TaobaoAd~\cite{Tianchi}, and KuaiVideo~\cite{li2019routing}, where the statistics are summarized in Table~\ref{tab:data}.

\begin{table*}[htbp]
\caption{Experiment results on real-world datasets. Methods with high gAUC and AUC, and low LogLoss are preferred.}
\label{tab:exp-result}
\begin{tabular}{l|lll|lll|lll}
\toprule
\multirow{2}{*}{\textbf{Method}} & \multicolumn{3}{c|}{\textbf{AmazonElectronics}}       & \multicolumn{3}{c|}{\textbf{TaobaoAds}}             & \multicolumn{3}{c}{\textbf{KuaiVideo}}           \\
                                 & \textbf{gAUC}   & \textbf{AUC}    & \textbf{LogLoss} & \textbf{gAUC} & \textbf{AUC} & \textbf{LogLoss} & \textbf{gAUC} & \textbf{AUC} & \textbf{LogLoss} \\ \midrule
xDeepFM     & 0.8770 & 0.8793 & 0.4382   & 0.5718 & 0.6402 & 0.1938  & 0.6622 & 0.7427 & 0.4371 \\
DCNv2                            &   0.8766        & 0.8791          &  0.4395          &  0.5735       &  0.6486      & 0.1930           &  0.6618       &  0.7420      &  0.4390               \\
DHEN                             &   0.8776        &   0.8799        &   0.4367         &   0.5708      &   0.6505     &   0.1929         &    0.6593     &   0.7428      &      0.4372            \\
Wukong &    0.8729     & 0.8767 &   0.4482   & 0.5656  & 0.6411  &  0.1935 & 0.6578  &  0.7417  & 0.4375 \\ \midrule
DIN                              &  0.8830         & 0.8858          &  {\ul 0.4319}         &  0.5752      &   0.6510     &  {\ul 0.1928}          &  \textbf{0.6657}       & 0.7440       & 0.4409         \\
DIEN                             &  0.8833         &  0.8860         &   0.4318        &  0.5731       &  0.6518      &  0.1932          &   0.6643            &    0.7444         &   0.4398                  \\ 
\rowcolor{cyan!20}
\modelname(DIN+DIEN)                           &  {\ul 0.8846}         &   {\ul 0.8869}       &   0.4346       &    0.5758     &   \textbf{0.6536}     &     \textbf{0.1922}     &        {\ul 0.6652}      &     0.7450        &     0.4367                \\ 
\midrule

InterFormer                             &  0.8832         & 0.8862        & 0.4316      & 0.5725        & 0.6520      & {\ul 0.1924}       &   0.6632            &  0.7445            &  {\ul 0.4356}            \\ 
\rowcolor{magenta!20}
\modelname(InterFormer+InterFormer)                             & 0.8842    & 0.8868 & {\ul 0.4297}  & {\ul 0.5762}        & {\ul 0.6530}       & 0.1931           &  0.6646       & {\ul 0.7453}        &  0.4360      \\ 
\rowcolor{magenta!20}
\modelname(InterFormer+DCNv2)                             &   0.8848  & 0.8870 & 0.4423  &    0.5724     &     0.6527   &    0.1921    &   0.6610     & 0.7428        &  0.4353    \\ \midrule
\midrule
\rowcolor{green!20}
\modelname(InteFormer+DHEN)                             & \textbf{0.8860}    & \textbf{0.8882} & \textbf{0.4247}  & \textbf{0.5767}        & \textbf{0.6536}       & 0.1930           &  0.6637       & \textbf{0.7456}        &  \textbf{0.4349}    \\
\bottomrule
\end{tabular}
\end{table*}

\begin{table}[htbp]
\caption{Dataset Summary.}
\label{tab:data}
\scalebox{0.93}{
\begin{tabular}{@{}llll@{}}
\toprule
Dataset           & \#Samples & \#Feat. (Seq/Global) & Seq Length \\ \midrule
Amazon            & 3.0M      & 6 (2/4)               & 100                            \\
TaobaoAd          & 25.0M     & 22 (3/19)             & 50                             \\
KuaiVideo         & 13.7M     & 9 (4/5)               & 100                            \\
\bottomrule
\end{tabular}}
\end{table}

\begin{table*}[htbp]
\caption{Ablation Study}
\label{tab:ablation}
\begin{tabular}{l|lll|lll|lll}
\toprule
\multirow{2}{*}{\textbf{Method}} & \multicolumn{3}{c|}{\textbf{AmazonElectronics}}       & \multicolumn{3}{c|}{\textbf{TaobaoAds}}             & \multicolumn{3}{c}{\textbf{KuaiVideo}}           \\
                                 & \textbf{gAUC}   & \textbf{AUC}    & \textbf{LogLoss} & \textbf{gAUC} & \textbf{AUC} & \textbf{LogLoss} & \textbf{gAUC} & \textbf{AUC} & \textbf{LogLoss} \\ \midrule
\rowcolor{green!20}
\modelname(InteFormer+DHEN)                             & 0.8860    & 0.8882 & 0.4247  & 0.5767   & 0.6536       & 0.1930           &  0.6637       & 0.7456        &  0.4349    \\ \midrule

\noindent w/o Confidence-based Fusion & 0.8833   & 0.8855 &  0.4287 &  0.5741  &   0.6508    &    0.1933    &  0.6628      &   0.7446     & 0.4355  \\

\noindent w/o KL Divergence &   0.8851 & 0.8878 & 0.4271  & 0.5712   &  0.6482     &  0.1939      &  0.6624      &   0.7443     & 0.4363  \\

\noindent w/o Multi-Embedding &  0.8851  & 0.8874 &  0.4290 &  0.5745  &   0.6512    &    0.1935    &   0.6621     &   0.7445     & 0.4357  \\

\noindent w/o Gradient Stop & 0.8834   & 0.8860 & 0.4296  &  0.5763  &   0.6532   &  0.1931     & 0.6620    &  0.7442    & 0.4377  \\
\midrule

\noindent Single-Embedding + concat &  0.8796  & 0.8828 & 0.4335  &  0.5699   &  0.6465     &    0.1934    &    0.6572   &   0.7389     &   0.4392 \\

\noindent Multi-Embedding + concat &  0.8780  & 0.8809 & 0.4360  & 0.5719   &  0.6495   &   0.1931   &   0.6545 &  0.7351  &  0.4413 \\

\bottomrule
\end{tabular}
\end{table*}

\subsubsection{Baseline Models.} 
To evaluate the effectiveness of our framework, we compare it against six state-of-the-art models. These include three non-sequential models—xDeepFM~\cite{xdeepfm}, DCNv2~\cite{dcnv2}, DHEN~\cite{DHEN}, and Wukong~\cite{Wukong}; and three sequential models-DIN~\cite{DIN}, DIEN~\cite{DIEN}, and InterFormer. 

\subsubsection{Hyperparameter Setting.}
Following the open-source CTR library~\footnote{\url{https://reczoo.github.io/FuxiCTR}}, we implement the baselines using their best-tuned hyperparameters and develop our model based on FuxiCTR~\cite{fuxiCTR}. All baselines achieve optimal performance with an embedding size of 64. In our framework, each component model operates with an embedding size of 32, while the original model configurations from the baselines remain unchanged. The only hyperparameters we tune are the learning rate, weight decay, and the weight of KL divergence $\alpha$, where $\alpha \in \{0.1, 0.5, 1.0\}$. 
To ensure reproducibility, we set the random seed to the commonly used value of 42.

\subsubsection{Evaluation Metrics.} 
We evaluate our model using three metrics: AUC, gAUC, and LogLoss. 
AUC (Area Under the Curve) measures the model’s ability to distinguish between positive and negative samples across all thresholds.
gAUC (Group AUC) evaluates ranking performance at the user level by computing AUC for each user and averaging them, providing insights into the model’s performance across different users.
LogLoss (Logarithmic Loss) measures the error between predicted probabilities and actual labels, with lower values indicating more accurate and reliable predictions.

\subsection{Performance Comparison}\label{sec:performance}

Table~\ref{tab:exp-result} summarizes the performance of our model compared to various baselines across three real-world datasets.
Higher gAUC and AUC values, and lower LogLoss values, indicate better performance.

Our results confirm that sequence-based models (e.g., DIN, DIEN) consistently outperform non-sequence models (e.g., DCNv2, DHEN, and WuKong) by a large margin. For instance, on the AmazonElectronics dataset, DIN achieves a 0.67\% improvement in AUC over DHEN, while DIEN improves AUC by 0.69\%. These results highlight the importance of modeling user behavior sequences for CTR tasks, as sequence-based models capture richer interaction patterns compared to non-sequence models.

Our framework effectively ensembles ("+") models to further enhance performance. Specifically, combining DIN and DIEN (DIN + DIEN) within our framework yields better results than either model individually. For example, on the TaobaoAds dataset, the ensemble improves AUC from 0.6518 to 0.6536 and reduces LogLoss from 0.1932 to 0.1922 compared with DIEN. 
This demonstrates that our framework enhances model complementarity, leading to improved ranking and prediction accuracy.
The best performance is achieved by combining Meta’s DHEN and InterFormer models using our framework. This combination achieves the highest gAUC, AUC, and lowest LogLoss across all datasets. 
On KuaiVideo, the DHEN+InterFormer ensemble improves AUC from 0.7428 to 0.7456 and reduces LogLoss by 0.53\% over DHEN. 
Similarly, on AmazonElectronics, it achieves a significant improvement in AUC from 0.8862 to 0.8882 over InterFormer and a 1.60\% reduction in LogLoss.

However, not all ensemble combinations yield the same level of improvement. When we ensemble InterFormer with itself or with DCNv2, the results are worse than those achieved by ensembling InterFormer with DHEN. For example, on the AmazonElectronics dataset, the ensemble of InterFormer and DCNv2 achieves a gAUC of 0.8848, which is slightly lower than the 0.8860 achieved by InterFormer+DHEN. A similar pattern is observed across other datasets, with the InterFormer+DHEN ensemble consistently achieving the best results. For instance, on KuaiVideo, InterFormer+DHEN improves AUC by 0.38\% compared to InterFormer+DCNv2 and achieves the lowest LogLoss of 0.4349.
We found that ensembling two identical models, such as InterFormer + InterFormer, also yields performance improvements. Further experiments on ensembling identical models are discussed in Case Study~\ref{ensemble case}.

\subsection{Ablation Study}\label{sec:Ablation}

To evaluate the contribution of each component in our model, we conduct an ablation study, as shown in Table~\ref{tab:ablation}. 
(1) Removing the \textbf{Confidence-based Fusion} module degrades performance across all datasets. For example, on AmazonElectronics, the gAUC drops from 0.8860 to 0.8833, indicating that adaptively weighting embeddings based on model confidence enhances prediction quality. 
(2) The absence of the \textbf{KL divergence} module also negatively impacts performance. For instance, on the KuaiVideo dataset, AUC decreases from 0.7456 to 0.7443, demonstrating that aligning the predictions of DHEN and InterFormer through KL divergence improves their collaboration.
(3) Similarly, the \textbf{Multi-Embedding module} is crucial for performance. When this component is removed, gAUC on TaobaoAds drops from 0.5767 to 0.5745, and LogLoss increases from 0.1930 to 0.1935. These results confirm that different embedding tables help capture diverse feature interaction patterns.
(4) Removing \textbf{Gradient Stop} allows gradients to flow through the confidence scores, causing instability in learning. As a result, performance slightly drops across datasets, such as an increase in gAUC from 0.6637 to 0.6620 on KuaiVideo, highlighting the importance of gradient stopping for stable fusion.

In the final two rows, we compare our multi-embedding design with simpler embedding schemes. Using a \textbf{Single Embedding table with concatenation} degrades performance significantly. For example, on AmazonElectronics, gAUC drops from 0.8860 to 0.8796, indicating that a single embedding table limits the ability to capture complex interactions.
Similarly, naively applying \textbf{Multi-Embedding with simple concatenation} at the output does not achieve the desired performance. For instance, AUC decreases from 0.6536 to 0.6495 on TaobaoAds. This suggests that merely concatenating embeddings from different tables does not fully leverage the complementary strengths of the models. In contrast, our fusion framework integrates the embeddings more effectively, balancing their contributions based on model confidence and alignment. 

\begin{figure}
     \centering
     \includegraphics[width=0.51\textwidth]{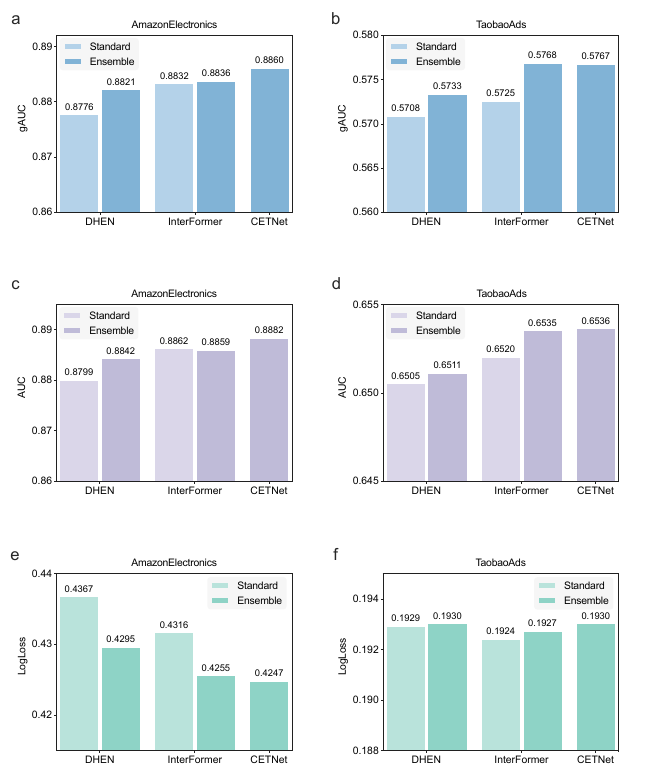}
    \caption{Performance of Component Model on AmazonElectronics and TaobaoAds}
    \label{case study}
\end{figure}

\begin{table*}[htbp]
\centering
\caption{Model performance on Criteo and Avazu datasets.}
\label{tab:mep}
\scalebox{0.96}{
\begin{tabular}{lllllll|lllll}
\toprule
\multicolumn{2}{c}{\multirow{2}{*}{Model}} & \multicolumn{5}{c}{Criteo} & \multicolumn{5}{c}{Avazu}  \\
\cmidrule(lr){3-7} \cmidrule(lr){8-12}
\multicolumn{2}{c}{} & base & 2x & 3x & 4x & 10x & base & 2x & 3x & 4x & 10x \\
\midrule
\multirow{4}{*}{DNN} &SE & \multirow{4}{*}{0.80915}  & 0.80948 & 0.80929 & 0.80949 & {\ul 0.80995} & \multirow{4}{*}{0.78747} & 0.78748 & 0.78742 & 0.78684 & 0.78589  \\
                      &ME &  & 0.80941 & 0.80952 & 0.80973 & 0.80978 &  & 0.78826 & 0.78829 & 0.78894 & {\ul 0.78894} \\
                      & Ours-sum &  & 0.80950 & {\ul 0.81010} & 0.81021 & 0.80989 &  & {\ul 0.78962} & \textbf{0.79000} & {\ul 0.78975} & 0.78582 \\ 
                    & ours-concat &  & {\ul 0.80960} & 0.80994 & \textbf{0.81030}  & 0.80959 &  & 0.78951 & 0.78947 & 0.78838 & 0.78843 \\ \midrule
\multirow{4}{*}{IPNN} &SE & \multirow{4}{*}{0.80936}  & 0.80914 & 0.80912 & 0.80916 & 0.80886  &\multirow{4}{*}{0.78713}  & 0.78778 & 0.78754 & 0.78722 & 0.78759 \\
                      &ME &  & 0.80855 & 0.80851 & 0.80864 & 0.80909 &  & 0.78815 & 0.78900 &  0.78923 & {\ul 0.78940} \\
                      & Ours-sum & & {\ul 0.80993} & 0.81020 & \textbf{0.81054} & {\ul 0.81041} &  & {\ul 0.78949} & \textbf{0.78979} & {\ul 0.78956}  &  0.78871\\ 
                    & Ours-concat &  & 0.80974 & {\ul 0.81035} & 0.81043 & 0.81020 &  & 0.78921 & 0.78934 &  0.78948 & 0.78885 \\ \midrule
\multirow{4}{*}{DCNv2} &SE & \multirow{4}{*}{0.81001}  & 0.81028  & 0.81038 & 0.81022 &  0.81019 & \multirow{4}{*}{0.78807} & 0.78854 & 0.78876 & 0.78890 & 0.78857 \\
                      &ME &   & 0.81028 & 0.81046 & 0.81046 & 0.81047 &  & 0.78880 & 0.78928  & 0.78905 & {\ul 0.78975} \\
                      & Ours-sum &  & 0.81061 & 0.81081 & {\ul 0.81098} & \textbf{0.81140} &  & 0.78967 & {\ul 0.79018} & 0.78962 & 0.78924  \\ 
                    & Ours-concat &  & {\ul 0.81069} & {\ul 0.81083} & 0.81087 & 0.81137 &  & \textbf{0.79033} & 0.79004 & {\ul 0.79007} & 0.78932 \\ \midrule
\multirow{4}{*}{FinalMLP} &SE & \multirow{4}{*}{0.80956}  & 0.80918 & 0.80982 & 0.80949 & 0.80986  &\multirow{4}{*}{0.78783}  & 0.78850 & 0.78803 & 0.78765 & 0.78615 \\
                      &ME &  & 0.80970 & 0.80945 & 0.80970 & {\ul 0.81008}  &  &  0.78821  & 0.78816  & 0.78802 & {\ul 0.78760} \\
                      & Ours-sum &  & {\ul 0.80995} & \textbf{0.81021} &  0.81015 & 0.81004 &  & 0.78948 & \textbf{0.78959} & {\ul 0.78901}  & 0.78730 \\ 
                    & Ours-concat &  & 0.80981 & 0.81012 & {\ul 0.81017} & 0.80995 &  & {\ul 0.78955} & \textbf{0.78959} & 0.78881 & 0.78751 \\ \midrule
\bottomrule
\end{tabular}}
\end{table*}

\subsection{In-depth Analysis}

\subsubsection{Evaluating Model Components}
Figure~\ref{case study} presents the performance of our ensemble framework and its component models on AmazonElectronics and TaobaoAds. The goal of this case study is to demonstrate that each individual model (DHEN and InterFormer) performs better within our ensemble framework than when trained independently.
Our results show that both DHEN and InterFormer benefit significantly from the collaborative learning and confidence-based fusion in our framework. 
For example, on the AmazonElectronics dataset, InterFormer alone achieves a gAUC of 0.8832, but when integrated into our ensemble with DHEN, the gAUC improves to 0.8836. 
Although the AUC of InterFormer decreases slightly from 0.8862 to 0.8859, the overall performance remains strong, with a notable improvement in LogLoss from 0.4316 to 0.4255.
Similarly, DHEN’s AUC improves from 0.8799 to 0.8842, a 0.49\% increase, when it participates in the ensemble framework.
A similar trend can be observed on the other datasets. 
On TaobaoAds, the AUC of InterFormer increases from 0.6520 to 0.6535, while the gAUC of DHEN improves from 0.5708 to 0.5733 within the ensemble. 


Overall, these results confirm that training DHEN and InterFormer as part of our ensemble framework results in better performance than training them independently. 

\subsubsection{Comparison with Multi-Embedding Paradigm}\label{ensemble case}

We evaluate the performance of our ensemble framework against the multi-embedding paradigm (ME) using the experimental setup outlined in \cite{multi-emb}. 
For a fair comparison, all experiments are conducted with consistent hyperparameters, including learning rate and weight decay. 
The base size is fixed at 10, and we investigate how varying the embedding table sizes and applying different fusion strategies influence performance on the Criteo and Avazu datasets, with the statistics of both datasets summarized in the Appendix (Table~\ref{tab: dataset2}).

Table~\ref{tab:mep} presents the results. Our ensemble framework is evaluated using two fusion strategies: \textbf{Ours-sum}, which sums the final embeddings, and \textbf{Ours-concat}, which concatenates them. These results are compared against both the single-embedding (SE) and multi-embedding (ME) baselines, where embedding sizes are scaled by 2x, 3x, 4x, and 10x.
On the Criteo dataset, FinalMLP with Ours-sum achieves the best AUC of 0.81021 with only a 3x embedding size. This result outperforms both the ME baseline (0.81008) and the SE model (0.80986) at the much larger 10x embedding size, demonstrating that our design can achieve efficient and high-quality representations with fewer parameters.
A similar trend is observed on the Avazu dataset. Although the ME framework with DCNv2 achieves an AUC of 0.78975 at a 10x embedding size, our framework with Ours-concat achieves a higher AUC of 0.79033 with just a 2x embedding size. 
This result highlights that our ensemble design not only matches but surpasses the performance of ME with far fewer embedding parameters. 

In industry, models are often trained on large datasets with only one epoch due to time and resource constraints.
We also conducted experiments to evaluate performance after one epoch of training, and our method once again demonstrated superior performance, as shown in Appendix~\ref{sec:app}.


\subsection{Internal Deployment}
\begin{figure}
     \centering     \includegraphics[width=0.45\textwidth]{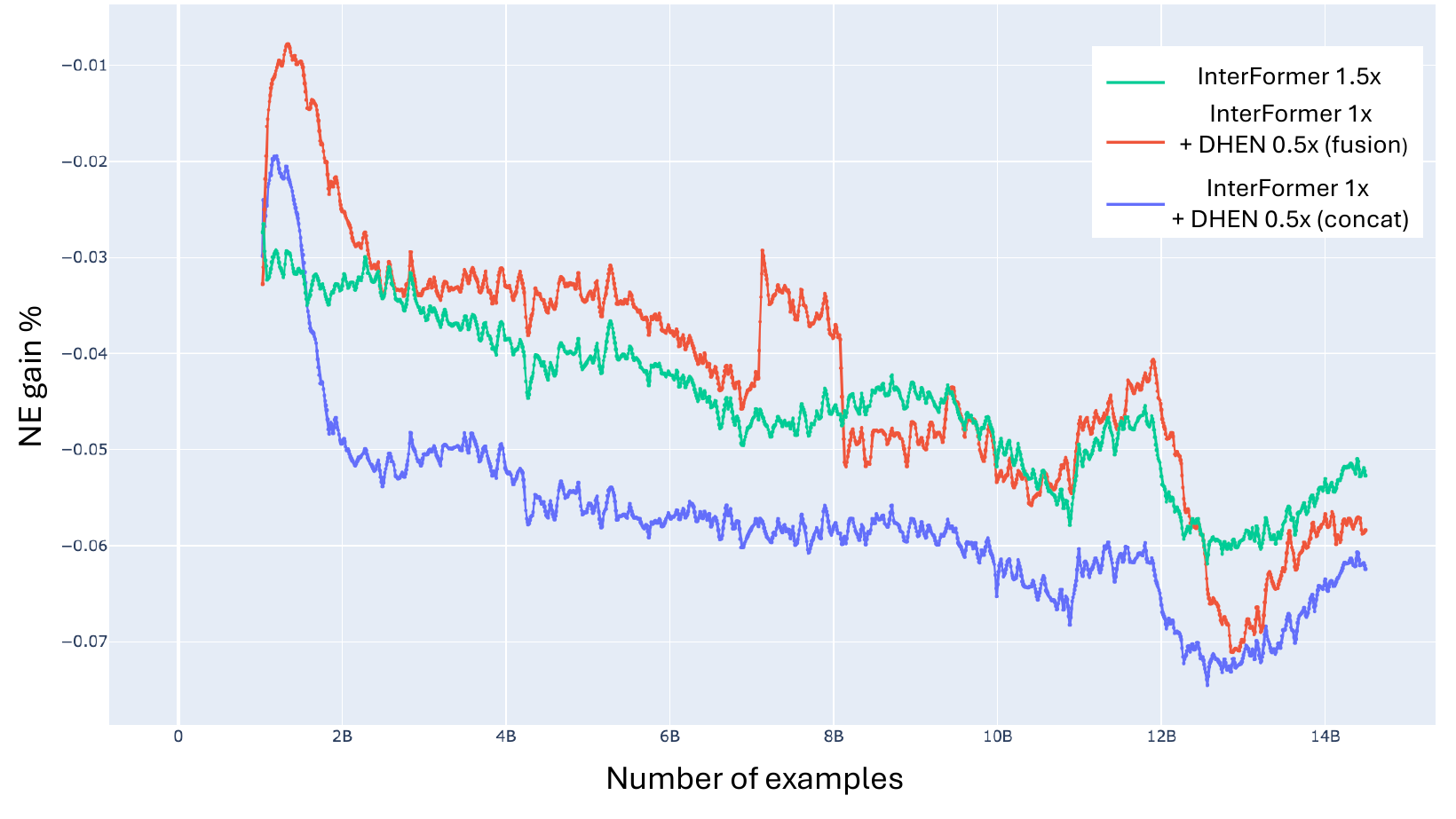}
    \caption{NE learning curve}
    \label{online}
\end{figure}
To evaluate the effectiveness of our framework in a real-world setting, we evaluate it on a large scale industrial dataset from Meta, using InterFormer with a fixed embedding size 
$d$ as the baseline. For comparative analysis, we conducted the following experiments: (1) InterFormer with an expanded embedding size of 1.5x, and (2) InterFormer with the original embedding size
$d$ enhanced by an additional 0.5x embedding table trained on a lightweight DHEN model, as shown in Figure~\ref{online}.

Normalized Entropy (NE)~\cite{NE} is the  optimization objective in Meta's ranking
systems.
Increasing the embedding size of InterFormer yields an NE gain of -0.051\%. 
In the second experiment, while confidence-based fusion does not result in significant improvements or stable learning in our internal setting, we find that simple concatenation, a specific configuration of the fusion mechanism, demonstrates greater stability and achieves a notably higher NE gain of -0.062\%. Although the difference may seem modest, this improvement is meaningful in our setting except in cases with a large number of examples, highlighting the effectiveness of a straightforward fusion approach for complex large-scale datasets.
Future work will explore intermediate-stage fusion within the multi-embedding paradigm to further improve performance.

\section{Conclusion}\label{sec:con}
In this paper, we address the limitations of multi-embedding paradigm by proposing a novel ensemble framework that leverages multiple models with distinct embedding tables to capture complementary interaction patterns. Through confidence-based fusion and collaborative learning with symmetric KL divergence, our framework dynamically balances model contributions and aligns predictions to prevent model dominance. Experiments on three public datasets, AmazonElectronics, TaobaoAds, and KuaiVideo, demonstrate the effectiveness of our approach, consistently outperforming individual models and state-of-the-art baselines. Further validation on Criteo and Avazu confirms that our framework achieves superior performance even with smaller embedding sizes, offering a more efficient and scalable solution for CTR prediction in both research and industrial applications.

\bibliographystyle{ACM-Reference-Format}
\bibliography{main}

\appendix
\begin{table*}[!t]
\centering
\caption{Model performance with one epoch training on Criteo and Avazu datasets.}
\label{tab:one_epoch}
\begin{tabular}{lllllll|lllll}
\toprule
\multicolumn{2}{c}{\multirow{2}{*}{Model}} & \multicolumn{5}{c}{Criteo} & \multicolumn{5}{c}{Avazu}  \\
\cmidrule(lr){3-7} \cmidrule(lr){8-12}
\multicolumn{2}{c}{} & base & 2x & 3x & 4x & 10x & base & 2x & 3x & 4x & 10x \\
\midrule
\multirow{4}{*}{DNN} &SE & \multirow{4}{*}{0.80465}  & 0.80423  & 0.80399 & 0.80373 & 0.80268  & \multirow{4}{*}{0.78162} & 0.78021 & 0.77969 & 0.77948 & 0.77704 \\
                      &ME &   & 0.80482 & 0.80480 & 0.80495 & {\ul 0.80468} &  & 0.78207 & 0.78187  & 0.78173 & 0.78162 \\
                      & Ours-sum &  & {\ul 0.80527} & \textbf{0.80542} & {\ul 0.80522}  & 0.80454 &  & \textbf{0.78353} & {\ul 0.78337} & {\ul 0.78334} &  0.78205 \\ 
                    & Ours-concat &  & 0.80524 & 0.80529 & 0.80510 & 0.80422 &  & 0.78347 & 0.78311 & 0.78318 & {\ul 0.78212} \\ \midrule
\multirow{4}{*}{IPNN} &SE & \multirow{4}{*}{0.80668}  & 0.80676 & 0.80695 & 0.80692 & 0.80599  &\multirow{4}{*}{0.78248}  & 0.78365 & 0.78266 & 0.78268 & 0.78238 \\
                      &ME &  & 0.80710 & {\ul 0.80727}  & {\ul 0.80736} & {\ul 0.80727} &  & 0.78404 & {\ul 0.78520} &  {\ul 0.78524} & \textbf{0.78544} \\
                      & Ours-sum & & \textbf{0.80745} & 0.80722 & 0.80701 & 0.80550 &  & 0.78492 & 0.78498 & 0.78461  & 0.78245 \\ 
                    & Ours-concat &  & 0.80731 & 0.80713 & 0.80692 & 0.80525 &  & {\ul 0.78522} & 0.78458 & 0.78493  & 0.78331 \\ \midrule
\multirow{4}{*}{DCNv2} &SE & \multirow{4}{*}{0.80742}  & 0.80763 & 0.80769 & 0.80745 & 0.80723 & \multirow{4}{*}{0.78377} & 0.78354 & 0.78362 & 0.78386 &  0.78348 \\
                      &ME &  & 0.80749  & 0.80749 & 0.80722 & 0.80685 &  & 0.78401 & {\ul 0.78428} & {\ul 0.78427} & \textbf{0.78464} \\
                      & Ours-sum &  & 0.80830 & {\ul 0.80808} & {\ul 0.80789} & {\ul 0.80727}  &  & {\ul 0.78453} & 0.78416 & 0.78358 & 0.78203 \\ 
                    & Ours-concat &  & \textbf{0.80835} & 0.80798 & 0.80777  & 0.80690 &  & 0.78432 & 0.78385 & 0.78383 & 0.78244 \\ \midrule
\multirow{4}{*}{FinalMLP} &SE & \multirow{4}{*}{0.78387}  & 0.80261 & 0.78372 & 0.80193 & 0.79018 &\multirow{4}{*}{0.78201}  & 0.74134 & 0.74056 &  0.78014 & 0.77817 \\
                      &ME &  & 0.71550 & 0.80049 & 0.80470 & 0.80112  &  & 0.78205  & 0.73633 &  0.78223 & 0.74126 \\
                      & Ours-sum &  & 0.80535 & {\ul 0.80514} & {\ul 0.80494}  & {\ul 0.80302} &  & 0.78353  & {\ul 0.78310} & 0.78267  & 0.78073 \\ 
                    & Ours-concat &  & \textbf{0.80538} & 0.80510 & 0.80467 & 0.80295 &  & \textbf{0.78400} & 0.78307 & {\ul 0.78294} & {\ul 0.78080} \\ \midrule
\bottomrule
\end{tabular}
\end{table*}
\newpage

\section{Appendix}\label{sec:app}

\subsection{Dataset}

\begin{table}[htbp]
\centering
\caption{The statistics of evaluation datasets.}
\label{tab: dataset2}
\begin{tabular}{l|r|r}
\toprule
\textbf{Dataset} & \textbf{\#Instances}  & \textbf{\#Features} \\
\midrule
Criteo      & 45,840,617  & 2,086,936 \\
Avazu       & 40,428,967  & 1,544,250 \\
\bottomrule
\end{tabular}
\end{table}

Table~\ref{tab: dataset2} provides detailed information about benchmark datasets:
\begin{itemize}
    \item \textbf{Criteo~\footnote{\url{https://www.kaggle.com/c/criteo-display-ad-challenge}}}: a well-known real-world benchmark for display advertising. It provides detailed information about individual ad displays along with corresponding user click feedback.
    \item \textbf{Avazu~\footnote{\url{https://www.kaggle.com/c/avazu-ctr-prediction}}}: contains several days of ad click-through data, arranged in chronological order. Each data entry includes 23 fields, representing various elements associated with a single ad impression.
\end{itemize}

\subsection{Experimental Results with One-Epoch Training}
To mimic conditions of industry setting, we conduct experiments on the Criteo and Avazu datasets, evaluating the performance of our ensemble framework under one-epoch training. Table~\ref{tab:one_epoch} presents the results, comparing our ensemble methods with the single-embedding (SE) and multi-embedding (ME) baselines.

On the Criteo Dataset, our ensemble framework consistently achieves the best performance when the embedding table size is small (2x or 3x). For example, with IPNN on Criteo, the Ours-sum strategy achieves the highest AUC of 0.80745 with 2x embedding size, outperforming both ME (0.80736) with 4x size and SE (0.80695) with 3x size. Similarly, with DNN, Ours-sum achieves the best AUC of 0.80542 with just a 3x embedding size, outperforming ME (0.80480) and SE (0.80399) models even with larger embedding tables. This demonstrates that our method is highly efficient and can reach optimal performance without relying on excessively large embeddings.
On the Avazu dataset, while ME achieves slightly better performance than our method in some cases, the difference is minimal, and our framework still offers significant advantages with smaller embedding tables. 
For instance, with DCNv2, ME attains an AUC of 0.78464 at a 10x size, while Ours-sum achieves an AUC of 0.78453 at a smaller 2x size.
This shows that even though the ME approach can marginally outperform ours in some settings, our method remains highly efficient, achieving competitive results with much smaller embedding sizes. 

\end{document}